\documentclass[conference]{IEEEtran}
\IEEEoverridecommandlockouts

\usepackage{cite}
\usepackage{amsmath,amssymb,amsfonts}
\usepackage{algorithmic}
\usepackage{graphicx}
\usepackage{textcomp}
\usepackage{xcolor, soul}

\usepackage{hyperref}
\usepackage[disable, colorinlistoftodos,prependcaption,textsize=tiny]{todonotes} %

\usepackage{pgfplots}
\usepackage{pgfplots}
\DeclareUnicodeCharacter{2212}{−}
\usepgfplotslibrary{groupplots,dateplot}
\usetikzlibrary{patterns,shapes.arrows}
\pgfplotsset{compat=newest}
\usepackage{hyperref}

\usepackage{makecell}
\usepackage{multirow}
\usepackage{xcolor}
\usepackage{svg}
\usepackage{amsmath}
\DeclareMathOperator*{\argmax}{arg\,max}

\usepackage{booktabs}
\usepackage{arydshln}

\let\svthefootnote\thefootnote
\newcommand\freefootnote[1]{%
  \let\thefootnote\relax%
  \footnotetext{#1}%
  \let\thefootnote\svthefootnote%
}
\usepackage{cite}

\newlength\figH
\newlength\figW
\pgfplotsset{/pgfplots/group/.cd,
    horizontal sep= 1.5cm,
    vertical sep= 1.5cm
}
\pgfplotsset{every tick label/.append style={font=\scriptsize}}

\def\BibTeX{{\rm B\kern-.05em{\sc i\kern-.025em b}\kern-.08em
    T\kern-.1667em\lower.7ex\hbox{E}\kern-.125emX}}

\begin{document}
\def\short_version{0} %
\def\all_tables{0} %

\title{persoDA: Personalized Data Augmentation for Personalized ASR}

\author{\IEEEauthorblockN{Pablo Peso Parada}
\IEEEauthorblockA{\textit{Samsung R\&D} \\
\textit{Institute UK (SRUK)} \\
United Kingdom \\
p.parada@samsung.com}
\and
\IEEEauthorblockN{Spyros Fontalis}
\IEEEauthorblockA{\textit{Centre for Research}\\
\textit{and Technology Hellas} \\
Greece \\}
\and
\IEEEauthorblockN{Md Asif Jalal}
\IEEEauthorblockA{\textit{Samsung R\&D} \\
\textit{Institute UK (SRUK)} \\
United Kingdom}
\and
\IEEEauthorblockN{Karthikeyan Saravanan}
\IEEEauthorblockA{\textit{Samsung R\&D} \\
\textit{Institute UK (SRUK)} \\
United Kingdom}
\and
\IEEEauthorblockN{Anastasios Drosou}
\IEEEauthorblockA{\textit{Centre for Research}\\
\textit{and Technology Hellas} \\
Greece \\}
\and
\IEEEauthorblockN{Mete Ozay}
\IEEEauthorblockA{\textit{Samsung R\&D} \\
\textit{Institute UK (SRUK)} \\
United Kingdom}
\and
\IEEEauthorblockN{Gil Ho Lee}
\IEEEauthorblockA{\textit{Language AI R\&D Group(MX)} \\
\textit{Samsung Electronics} \\
South Korea}
\and
\IEEEauthorblockN{Jungin Lee}
\IEEEauthorblockA{\textit{Language AI R\&D Group(MX)} \\
\textit{Samsung Electronics} \\
South Korea}
\and
\IEEEauthorblockN{Seokyeong Jung}
\IEEEauthorblockA{\textit{Language AI R\&D Group(MX)} \\
\textit{Samsung Electronics} \\
South Korea}
}

\makeatletter
\def\@IEEEmaketitle{\vskip -100pt\@maketitle}
\makeatother

\maketitle

\begin{abstract}
Data augmentation (DA) is ubiquitously used in training of Automatic Speech Recognition (ASR) models. DA offers increased data variability, robustness and generalization against different acoustic distortions. Recently, personalization of ASR models on mobile devices has been shown to improve Word Error Rate (WER). This paper evaluates data augmentation in this context and proposes \textit{persoDA}; a DA method driven by user's data utilized to personalize ASR  \cite{munkhdalai2022fast,sim21_interspeech,UserLibri}. {\it persoDA} aims to augment training with data specifically tuned towards  acoustic characteristics of the end-user, as opposed to standard augmentation based on Multi-Condition Training (MCT) that applies random reverberation and noises. Our evaluation with an ASR conformer-based baseline trained on Librispeech and personalized for VOICES\cite{voices} shows that {\it persoDA} achieves a 13.9\% relative WER reduction over using standard data augmentation (using random noise \& reverberation). Furthermore, {\it persoDA} shows 16\% to 20\% faster convergence over MCT.
\end{abstract}


\section{Introduction} %
End-to-end (E2E) Automatic Speech Recognition (ASR) models achieve state-of-the-art results \cite{li2022recent} when trained with thousands of hours of speech data from multiple users. However, one of the challenges of on-device deployment of these ASR models arises from the mismatch between training and test conditions; these models suffer from performance degradation due to speech variability \cite{benzeghiba2007automatic} or different acoustic environments (such as additive noise or reverberation) of the target user. Personalization of ASR models aims to reduce this performance gap by leveraging limited user data.

A variety of techniques have been employed to boost the personalization task, such as contextual biasing \cite{munkhdalai2022fast}, embedding based adaptation \cite{bell2020adaptation, zeineldeen2022improving} or feature transformation adaptation \cite{liu2021recent}. However, techniques proposed for using personalized data augmentation (DA) in ASR tasks have been scarce. %
A study of personalization of deep learning models \cite{schneider2021personalization} for computer vision tasks investigates methods used to augment user data by adding similar data identified via  auto-encoders. %
Employment of neural language generator and personalized speech synthesis to synthesize personalized speech for rapid speaker adaptation is investigated for hybrid systems \cite{huang2020using} and RNN-T \cite{huang2020rapid} models. An informed method of enabling automatic selection and parameterization of a DA pipeline for self-supervised learning is introduced in \cite{zaiem2022automatic}. 
A method to select a subset of the augmented data is proposed in \cite{Doulaty2016} using cosine similarity over the phoneme's posteriors, which is applied to the domain mismatch and  not to the personalization.

\begin{figure}[t]
  \centering
  \includegraphics[width=0.83\linewidth, page=5,trim={3cm 2.9cm 15.7cm 10.9cm},clip]{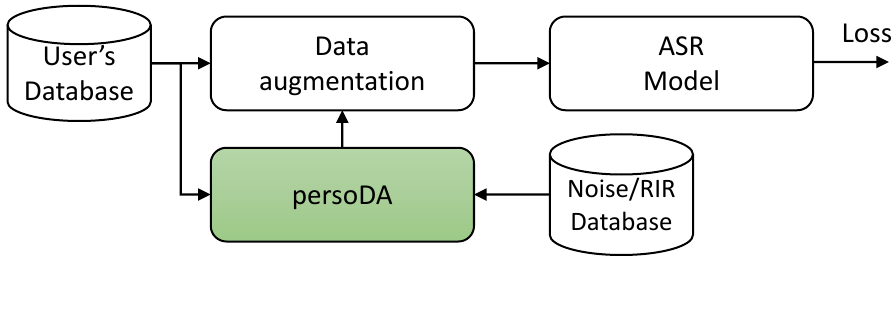}\vspace*{-20pt}
  \caption{The personalized data augmentation ({\it persoDA}) framework. {\it persoDA} guides the training DA process to select the most adequate augmentation given user's data.}\vspace*{-15pt}
  \label{fig:persoDA}
\end{figure}

In \cite{jin2022personalized}, DA methods  
based on  speaker dependent generative adversarial networks (GAN) are proposed for atypical speech, namely dysarthric and elderly, to increase the training data. 
In \cite{sivaraman2021personalized}, a data purification approach is proposed to  improve the training of a self-supervised personalized speech-enhancement model. %
Several DA methods have been proposed to improve ASR performance such as SpecAugment \cite{park2019specaugment}, Multi-Condition Training (MCT) \cite{rir_datasets}, patched Multi-Condition Training (pMCT) \cite{pesoparada22_interspeech}, and G(raph)-Augment \cite{wang2023g}.
However, none of them target ASR personalization tasks. %

In this work, we present a novel method to perform personalized data augmentation for an on-device ASR personalization task. Our contributions are:
\begin{itemize}
\item We propose a framework called {\it persoDA} (illustrated in Fig.~\ref{fig:persoDA}) to perform personalized data augmentation with faster convergence, given limited unlabelled user data.
\item Our experimental results show the effectiveness of {\it persoDA} for two specific cases of extracting information from unlabeled user data to perform (1) personalized noise data augmentation ({\it persoNoise}), and (2) personalized room acoustic data augmentation ({\it persoReverb}).
\end{itemize}

\section{Method}\label{sec:method}
In our personalization task, ASR models are required to be adapted to improve their performance on target user data.
Our method ({\it persoDA}) presented in Fig.~\ref{fig:persoDA} is based on extracting prior information from data distribution $p$ (defined by the user's recordings) about noise and reverberation which is available in the support $\sigma{(p)}$ to guide the personalized data augmentation.
This enables adaptation of the ASR model to acoustic conditions observed in a user's recordings.
This problem can be formulated as optimizing data augmentation parameters $\theta$ that maximize the probability $P(\theta \,|\, E)$ of generating similar acoustic environments by
\begin{equation}%
    \argmax_{\theta} P(\theta \,|\, E)
    \label{eq:argmax}
\end{equation}
where $E$ denotes the acoustic environment characteristics observed in the data.

The most common signal distortions present in speech recordings are additive noise, due to the presence of background sound sources, and reverberation, due to the fact of generating speech in an enclosed space. These two types of augmentations are considered in this work. The methods proposed for personalizing models to these audio distortions are called {\it persoNoise} and {\it persoReverb}, respectively.

The personalization task is performed on the user's recording set $\mathcal{U}$ which is split into two mutually exclusive subsets: (i) a subset $\mathcal{V}$ containing recordings with a high level of acoustic distortions (such as noise or reverberation); (ii) a subset $\mathcal{T}$ containing those with low levels of distortions. 
This subset can be identified by applying signal-to-noise estimators such as \cite{wada} and T60 estimators such as \cite{lollmann2010improved} to select the clean utterances for the set. 
The subset $\mathcal{T}$ is used as training samples utilized for model personalization.
Therefore, the data augmentation methods are applied only to this subset, and $\mathcal{V}$  or a subset of $\mathcal{V}$ is used by {\it persoDA} to optimize $\theta$ in \eqref{eq:argmax}.
Note that $\mathcal{U} = \mathcal{T} \cup \mathcal{V}$, and $\mathcal{T} \cap \mathcal{V} = \emptyset$.

\subsection{persoNoise}\label{personoise_method}
Here, persoNoise follows our hypothesis that specific noise characteristics of each user provides better augmentation as compared to standard and generic augmentation methods such as based on MCT. Thus, our methodology extracts noise present in $\mathcal{V}$ to be used for augmentation with our ASR personalization task involving user set $\mathcal{T}$. Therefore, the parameterization of $\theta$ in (\ref{eq:argmax}) is estimated by optimizing (\ref{eq:argmax})  on a set of persoNoise augmented recordings.

Two different methods, i) Voice Activity Detection (VAD) based and ii) Sound Source Separation based methods, are proposed to extract the background noise from recordings. In both cases, the acoustic environment characteristics observed in the data $E$ is the background noise present in the speech and $\theta$ is the best estimate of the noise.
As explained in the following sections,  the parameter $\theta$ is estimated in the VAD based approach from the non-speech audio segments whereas for the later approach the estimation of $\theta$ is based on separating the audio into different sources and select the source that is most likely to be background noise.

\subsubsection{VAD based Noise Extraction}\label{sec:VAD}
The proposed noise extraction method is based on a Python implementation of WebRTC VAD\cite{pywebrtcvad}. The VAD module classifies every input frame into speech or non-speech. In this work, VAD is used to select the {\it noise} (non-speech) frames  ${\mathbf{z}_m=[z_1, z_2, \cdots, z_n]}$, $n\geq1$, where $\mathbf{z}_m \in \mathbf{R}^n$ is the $m^{th}$ frame. %

The consecutive noise frames detected by the VAD module are concatenated to create noise segments $\mathbf{s}_o=[\mathbf{z}_m,\mathbf{z}_{m+1}, \cdots, \mathbf{z}_{m+p}]$, $\forall o,m,p\geq1$  and normalized in amplitude. The employed normalization method is based on modifying the amplitude of the noise segment such that the Root Mean Square (RMS) level is 5dB.
This normalization ensures that the noise segments contain similar levels, so that they can be concatenated smoothly. The noise segments longer than a threshold are concatenated to generate the noise recordings ${\mathbf{r}}=[\mathbf{s}_1, \mathbf{s}_2, \cdots, \mathbf{s}_q],\, q\geq1$  which are used to create the {\it persoNoise} dataset. 

The noise segments are randomly concatenated following a uniform distribution. A linear crossfade is applied in the boundaries of the concatenated segments to avoid large discontinuities in the signal. %
The concatenation stops when the length of the generated noise recording is longer than the longest user utterance in the training set.

Additionally, similar noises can be retrieved from a generic noise dataset to be included in the {\it persoNoise} dataset. %
This technique constitutes a method of data grouping \cite{schneider2021personalization}, which is defined as augmenting the data of an individual by adding similar samples taken from another broader noise dataset $\mathcal{B}$.  The similarity between a noise segment $\mathbf{s}_q$ and another noise segment $\mathbf{\hat{s}}$ extracted from a sample $b \in \mathcal{B}$ is computed using the cosine similarity in the power spectral density domain. %

Thereby, the concatenated noise segments $\mathbf{s}_q$ of $\mathbf{r}$ serve as audio spectral reference of the noises encountered in each individual's recordings. Thus, for each ensemble noise segment  $\mathbf{s}_q$, the most similar noise sample of an external noise dataset is picked and added in the {\it persoNoise} dataset.

\subsubsection{Speech Separation based Noise Extraction}\label{sec:SSS}
An alternative to extracting the noise profile from non-speech segments, a speech separation model can be employed to extract the background noise \cite{sepformer}.
The separation module used is the SepFormer model \cite{sepformer}. Once the audio source in a recording is separated, the noise channels are selected.
 This selection is carried out using the VAD and saving the channel with the least amount of speech segments in the {\it persoNoise} dataset.

\vspace*{-2pt}
\subsection{persoReverb}\label{sec:reverb}
Similar to persoNoise, persoReverb extracts reverberation present in $\mathcal{V}$ to be used for augmentation with ASR personalization task involving user set $\mathcal{T}$. T60 \cite{kuttruff2016room} is used to measure level of reverberation, where T60 is defined as the time needed for the sound pressure level in the room to drop 60 dB after the acoustic excitation ceases~\cite{kuttruff2016room}. Thus, $\theta$ in (\ref{eq:argmax}) represents an estimation of T60 from the reverberant acoustic environment $E$ observed in the input data.

 In this work, two different methods are evaluated to measure the level of reverberation. The first one is based on a statistical modelling of the sound decay \cite{lollmann2010improved}.
 The second estimation method is an algorithm based on spectral decay distributions \cite{eaton2013noise}. Once the level of reverberation is estimated for a given utterance, the RIR from the database with the closest T60 is drawn from the RIR database and stored in the {\it persoReverb} dataset.

\section{Evaluation}\label{sec:eval}

\subsection{Experimental Setup}
We implemented the ASR baseline using the transformer recipe in SpeechBrain \cite{speechbrain} for LibriSpeech.
The recipe implements an end-to-end transformer ASR architecture with a Conformer encoder \cite{conformer}.
The Conformer configuration follows the Conformer (S) described in \cite{conformer}.

In the personalization experiments, we fine-tune all the parameters of the pre-trained ASR models with Adam optimizer and a fixed learning rate of 2.5e-5. SpecAugment is also applied following the LibriFullAdapt policy. The {\it persoDA} method is compared to the MCT augmentation \cite{rir_datasets} (named as + reveberation and + noise in Section \ref{sec:analysis}) where the noise and RIR  are randomly selected and added. The best probability for MCT of including reverb or including noise is found empirically for each method.

\subsection{Datasets}
{\bf Pre-trained model dataset}: The ASR model is trained on LibriSpeech \cite{librispeech} using MCT  achieving 4.13\% WER in dev-clean. %
The LibriSpeech dataset comprises 960h of English audiobook utterances. %
The total size of the training data is reduced to 940h after excluding the speakers included in validation set $\mathcal{V}$ to avoid data overlap with the evaluation set of the VOiCES dataset \cite{voices}.

{\bf Personalization datasets}: Three different sets are created to perform the personalization experiments:
\begin{itemize}%
\item The global set, used as a measure of {\it catastrophic forgetting}\cite{catastrophicforgetting}, is %
created from a 14 minutes random subset from LibriSpeech dev-clean to accelerate experiments.
\item Validation set $\mathcal{V}$ is designed using data of 18 users from the VOiCES  evaluation set containing more utterances. VOiCES is created by recording LibriSpeech utterances played in acoustically challenging conditions. This dataset allows assessing accuracy of the proposed methods in realistic and noisy reverberant environments.
\item The personalization training set $\mathcal{T}$ is used to adapt the pre-trained ASR model. It consists of clean versions of the noisy or reverberant utterances present in $\mathcal{V}$. That is, $\mathcal{V}_{i,j}=f_j\left(\mathcal{T}_i\right)$ where $f_j$ is the augmentation $f(\cdot)$ created by replaying recording $\mathcal{T}_i$ in a noisy room.
\end{itemize}

On average, the duration of $\mathcal{T}$ per speaker is 3.41~min (with 13 utterances) and 12.94~min (with 50 utterances) for $\mathcal{V}$. %

{\bf Datasets used for DA}: Two main sets are used for DA: (1)
comprising RIRs, and (2) with noise recordings. We use a set
of RIRs gathered from 3 datasets \cite{rir_datasets}, giving rise to 325 different RIRs. The noise dataset contains 843 noise samples obtained from the Freesound portion of the MUSAN \cite{musan}. The SNR is sampled from a uniform distribution between 0-30 dB.

\subsection{Evaluation Metrics}
The performance metric used is Word Error Rate (WER). The Global WER indicates the level of catastrophic forgetting and Valid WER, computed on $\mathcal{V}$ or subset of $\mathcal{V}$, shows the personalization performance. The personalized WER is computed across different personalized models on each user's validation set.

\section{Experimental Analyses}\label{sec:analysis}
In this section, we analyse the performance of the proposed methods compared to standard or generic DA methods (MCT). Noise and reverberation characteristics for {\it persoNoise} and {\it persoReverb}, parametrized by $\theta$, are obtained from $\mathcal{V}$, and randomly applied to the personalization task involving utterances in $\mathcal{T}$. In sections \ref{sec:results_persoNoise} to \ref{sec:results_persoDA}, the distribution of $\theta$ applied to $\mathcal{T}$ overlaps with that found in $\mathcal{V}$. Whereas, as an ablation, we also study the impact of having disjoint distributions in section \ref{sec:PL}.

\if\short_version0
\subsection{Analysis of \textit{persoNoise}\label{sec:results_persoNoise}}
Table \ref{tab:results_persoNoiseReverb} shows the comparison of the three {\it persoNoise} techniques presented in Section \ref{personoise_method}.
The crossfade used for all experiments is 100~ms.
The best performance is achieved by the VAD based  noise extraction method without data grouping.
This VAD based approach is computationally lighter compared to the speech separation based approach (Section \ref{sec:SSS}). The WebRTC VAD algorithm uses a Gaussian Mixture Model to perform the speech-noise detection, whereas the speech separation can only be performed effectively by large deep learning models (in this case, with 25.7M parameters). Thus, {\it persoNoise} refers to the VAD based approach without grouping in the remaining of the paper.

\subsection{Analysis of \textit{persoReverb}} 
The performance comparison of the two methods for T60 estimation is shown in Table \ref{tab:results_persoNoiseReverb}. The table suggests that the sound decay modelling provides lower WER. Therefore, {\it persoReverb} used in the following analyses refers to the sound decay statistical modeling based T60 estimation method presented in Section \ref{sec:reverb}.

\vspace*{-10pt}\begin{table}[h]          
	\caption{Performance in terms of WER (\%) of different generation techniques. Trained for 100 steps.}\vspace*{-8pt}
	\label{tab:results_persoNoiseReverb}
	\centering
	\begin{tabular}{crcc}
		\toprule
		\multicolumn{2}{c}{\multirow{2}{*}{{\bf Method}}}  & \multicolumn{2}{c}{\bf Evaluation Set}  \\ 
		&  &  \makecell{Global} & \makecell{Valid}\\
		\midrule
		\multirow{2}{*}{{\it persoNoise}} & VAD based & 4.25 & {\bf 14.58} \\
		& VAD based + Data grouping  & {\bf 4.2} & 14.74 \\
		& Speech separation based & 4.24 & 15.1 \\
		\midrule
		\multirow{2}{*}{{\it persoReverb}} & Sound decay modeling based & 4.3 & {\bf 11.53} \\
		& Spectral decay distributions based & {\bf 4.26} & 11.69 \\
		\bottomrule
	\end{tabular}\vspace*{-5pt}
\end{table}

\fi

\subsection{Comparison of \textit{persoDA}\label{sec:results_persoDA}}
Table \ref{tab:results} provides a comparison for different combinations of the personalized and non personalized DA methods, when models are trained for 1000 steps. The performance of the proposed approach (+ {\it persoReverb} + {\it persoNoise}) achieves the best performance overall, reducing the WER by 13.88\% compared to random selection of noises and RIRs. 

Baseline personalization experiment is based on simply fine-tunning models on the training data without MCT augmentation. Table \ref{tab:results} also indicates that the personalized data augmentations outperform the random data augmentation. Indeed, {\it persoReverb} reduces the error by 7.45\% relative WER compared to the random counterpart, and {\it persoNoise} decreases the error by 10.81\% relative WER, which proves our initial hypothesis (Section \ref{personoise_method}) that specific noise characteristics  of  each  user provides better augmentation.

\begin{table}[ht]\vspace*{-10pt}
  \caption{WER (\%) of models trained for 1000 steps. Methods above the dashed line do not augment data. }\vspace*{-8pt}
  \label{tab:results}
  \centering
  \begin{tabular}{rcc}
    \toprule
    \multirow{2}{*}{{\bf Method}} & \multicolumn{2}{c}{\bf Evaluation set}  \\ 
     &  \makecell{Global} & \makecell{Valid}\\
    \midrule
    No personalization & 4.13 & 18.22\\
    Baseline personalization & 4.78 & 7.15 \\
    \hdashline
    + reverberation & 4.71 & 3.89 \\
    + noise & 4.72 & 7.31 \\
    + reverberation + noise & 4.68 & 3.53  \\
    + {\it persoReverb} & 4.76 & 3.6 \\
    + {\it persoNoise} & 4.8 & 6.52 \\
    + {\it persoReverb} + noises & 4.73 & 3.35 \\
    + reverberation + {\it persoNoise} & 4.69 & 3.26 \\
    + {\it persoReverb} + {\it persoNoise} & {\bf 4.68} &{\bf 3.04 } \\ 
    \bottomrule
  \end{tabular}%
\end{table}

\textbf{An ablation study for training models with different number of training steps.} Table \ref{tab:results_steps} presents a comparison of WER of models trained using reverberation and noise against using {\it persoReverb} and {\it persoNoise} with different number of training steps. The results indicate that after 100 steps, models achieve a relative WER improvement of 4.79\% in the validation set and slight improvement in the global set. This relative improvement obtained using the validation set keeps increasing with the number of steps reaching a relative WER improvement of 13.88\% without degradation on the global set.
\addtolength{\tabcolsep}{-2pt} %
\begin{table}[t]
  \caption{Comparison of standard DA (MCT) and {\it persoDA}. The  displayed values are Valid WER (\%) / Global WER (\%). Rel. Impr. row shows relative improvement of {\it persoDA} over MCT.}\vspace*{-8pt}
  \label{tab:results_steps}
  \centering
  \begin{tabular}{rcccc}
    \toprule
    \multirow{2}{*}{\bf Method} & \multicolumn{4}{c}{\bf Number of steps} \\
  	& 100 & 250 & 500 & 1000 \\
    \midrule
	MCT &11.89/4.26 & 7.86/4.35 & 5.31/4.52 & 3.53/4.68\\
    {\it persoDA}&11.32/4.23 & 7.38/4.35 & 4.89/4.54 & 3.04/4.68 \\
	\hdashline
	Rel. Impr.&4.79/0.70 & 6.11/0.00 & 7.91/-0.44 & 13.88/0.00\\	
    \bottomrule
  \end{tabular}\vspace*{-0 pt}
\end{table}

\vspace*{-15pt}\subsection{Evaluation on disjoint training and validation sets with and without pseudo-labels \label{sec:PL}} 
In this experiment, the training set $\mathcal{T}$ is split into 2 subsets with equal number of utterances: $\mathcal{T}'$ and $\mathcal{T}''$. A similar split is applied to $\mathcal{V}$ where $\mathcal{V}'$ and $\mathcal{V}''$ comprise the noisy reverberant versions of the utterances in $\mathcal{T}'$ and $\mathcal{T}''$ respectively. Thus, in this setup, the training phase of model personalization uses only $\mathcal{T}'$ set, and {\it persoDA} uses only $\mathcal{V}'$ to estimate information to guide the DA. This allows to measure the performance of models on the unseen acoustic environments in $\mathcal{V}''$. In addition, we assess the performance of the proposed methods with ground truth labels and with pseudo-labels \cite{PL} to simulate scenarios where the end user is not providing feedback on the hypothesis generated by the ASR.

Fig.~\ref{fig:results_train} shows the  performance on $\mathcal{V}'$ indicating that the proposed approach outperforms the MCT approach w/ labels and w/o labels (i.e. using pseudo-labels).
The graph also suggests faster converge for {\it persoDA}. Indeed, the final WER achieved by MCT w/ labels and w/o is reached 80 and 100 steps earlier respectively with the {\it persoDA} counterparts, thus, achieving 16\% to 20\% faster convergence. %
\begin{figure}[ht!]\vspace*{-10pt}
  \begin{tikzpicture}
    \begin{axis}[grid=both,
                 grid style={solid,gray!30!white},
                 minor y tick num=5,
                 no markers,
                 mark size=1.1pt,
                 height=40mm,
                 ymin=3.5,ymax=15,
                 xmin=0,xmax=500,
                 smooth, 
                 width=\linewidth,
                 xlabel={Steps},
                 ylabel={WER (\%)},
                 x label style={at={(axis description cs:0.5,-0.065)},anchor=north, font=\footnotesize},
                 y label style={at={(axis description cs:-0.065,.5)},rotate=0,anchor=south, font=\footnotesize},
                 every axis plot/.append style={thick},
                 legend columns=2, 
                 legend style={row sep=0.001 pt, font=\scriptsize, at={(1.1,1)},anchor=north east},
                 tick label style={font=\footnotesize} ]
      \addplot table[smooth,x index=0,y index=1,col sep=comma] {./data_disjoint_train.csv};
      \addplot table[x index=0,y index=2,col sep=comma] {data_disjoint_train.csv};
      \addplot table[x index=0,y index=1,col sep=comma] {data_disjoint_pl_train.csv};
      \addplot table[x index=0,y index=2,col sep=comma] {data_disjoint_pl_train.csv};
      \addlegendentry{MCT w/ labels}
      \addlegendentry{persoDA w/ labels}
      \addlegendentry{MCT w/o labels}
      \addlegendentry{persoDA w/o labels}
    \end{axis}
  \end{tikzpicture}\vspace*{-10pt}
  \caption{WER achieved on the set $\mathcal{V}'$ used to estimate information to guide {\it persoDA}. Experiments `w/ labels' were trained with ground-truth transcripts and `w/o labels' were trained with pseudo-labels.}\vspace*{-15pt}
\label{fig:results_train}
\end{figure}
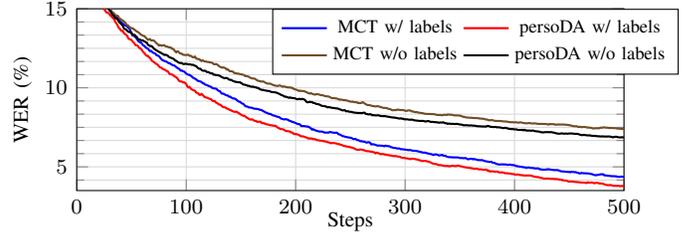

\begin{figure}[ht!]
  \begin{tikzpicture}
    \begin{axis}[grid=both,
                 grid style={solid,gray!30!white},
                 minor y tick num=5,
                 no markers,
                 height=40mm,
                 width=\linewidth,
                 xlabel={Steps},
                 xmin=0,xmax=500,
                 ymin=14.8,ymax=16.5,
                 smooth, 
                 ylabel={WER (\%)},
                 x label style={at={(axis description cs:0.5,-0.065)},anchor=north, font=\footnotesize},
                 y label style={at={(axis description cs:-0.1,.5)},rotate=0,anchor=south, font=\footnotesize},
                 legend columns=2, 
                 legend style={font=\scriptsize},
                 every axis plot/.append style={thick},
                 tick label style={font=\footnotesize} ]
      \addplot table[x index=0,y index=1,col sep=comma] {data_disjoint_valid.csv};
      \addplot table[x index=0,y index=2,col sep=comma] {data_disjoint_valid.csv};
      \addplot table[x index=0,y index=1,col sep=comma] {data_disjoint_pl_valid.csv};
      \addplot table[x index=0,y index=2,col sep=comma] {data_disjoint_pl_valid.csv};
      \addlegendentry{MCT w/ labels}
      \addlegendentry{persoDA w/ labels}
      \addlegendentry{MCT w/o labels}
      \addlegendentry{persoDA w/o labels}
    \end{axis}
  \end{tikzpicture}\vspace*{-10pt}
  \caption{WER achieved on the set $\mathcal{V}''$ which comprises unseen data. Experiments `w/ labels' were trained with ground-truth transcripts and `w/o labels' were trained with pseudo-labels.}\vspace*{-5pt}
\label{fig:results_valid}
\end{figure}
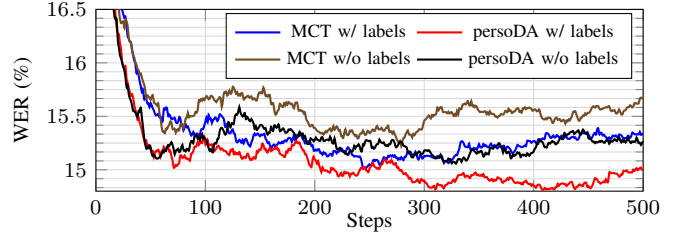

Fig.~\ref{fig:results_valid} displays the WER achieved on the $\mathcal{V}''$ set, i.e. the set with unseen acoustic conditions, which indicates that {\it persoDA} achieves lower WER compared to the MCT approach w/ and w/o labels. The plot also suggests that the best performance when using pseudo-labels is rapidly achieved (before 100 steps) without improving significantly as the steps increase. We hypothesize that this is due to the errors on the pseudo-labels (the WER is 3.71\%) and it could be further improved by generating new pseudo-labels as in \cite{slimIPL}.

\section{Conclusion}\label{sec:conclusions}
We have introduced a novel personalized data augmentation framework called {\it persoDA} to adapt ASR models to users' acoustic characteristics. This framework takes into consideration two types of acoustic distortions present in on-device recordings, background noise and reverberation, and personalizes  the corresponding sets for specific users. The proposed approach has low complexity and requires only a set of RIRs (typically of only 1 second long each) to be stored on the target platform, as opposed to MCT where RIRs and long noise recordings need to be saved or methods as \cite{Doulaty2016} where a exponential search is required.
Experimental analyses show the effectiveness of the {\it persoNoise} and {\it persoReverb} for ASR personalization tasks, reducing the error by 13.88\% relative WER compared to the standard DA methods, which use random noises and RIRs \cite{rir_datasets}.

\bibliographystyle{IEEEtran}

\bibliography{mybib}

\end{document}